\documentstyle[11pt,newpasp,twoside,epsf]{article}
\def\edcomment#1{\iffalse\marginpar{\raggedright\sl#1\/}\else\relax\fi}
\reversemarginpar

\begin{document}
\title{Spectra and power of blazar jets}
\author{Gabriele Ghisellini}
\affil{Osservatorio Astron. di Brera, Via Bianchi 46, Merate, I--23807, Italy}

\begin{abstract}
The spectra of blazars form a sequence which can be parame\-trized in term
of their observed bolometric luminosity.
At the most powerful extreme of the sequence we find objects whose
jet power can rival the power extracted by accretion, while
at the low power end of the sequence we find TeV and TeV candidate
blazars, whose spectral properties can give information on the particle
acceleration mechanism and can help to measure the IR background.
Most of the emission of blazars is produced at a distance of few hundreds
Schwarzschild radii from the center, and must be a small fraction of 
the kinetic power carried by the jet itself: most of the jet energy must be 
transported outwards, to power the extended radio structures.
The radiation produced by jets can be the result of internal shocks 
between shells of plasma with different bulk Lorentz factors.
This mechanism, thought to be at the origin of the gamma--rays observed in
gamma--ray bursts, can work even better in blazars, explaining
the observed main characteristics of these objects.
{\it Chandra} very recently discovered large scale X--ray jets both in
blazars and in radio--galaxies, which can be explained 
by enhanced Compton emission with respect to a pure 
synchrotron self Compton model.
In fact, besides the local synchrotron emission,
the radiation coming from the sub--pc jet core and the cosmic
background radiation can provide seed photons for the scattering process,
enahancing the large scale jet X--ray emission in 
radio--galaxies and blazars, respectively.
\end{abstract}

\section{Introduction}

The discovery that blazars are strong $\gamma$--ray emitters
is certainly one of the great achievements of the {\it Compton
Gamma Ray Observatory} mission (e.g. Hartmann et al., 1999)
and the ground based Cherenkov telescopes 
(e.g. Weekes et al. 1996; Petry et al. 1996)
This has allowed to study relativistic jets knowing, 
at last, the total amont of radiative power produced 
by them, and at which frequencies their spectra peak.
But also X--ray observations have been, and will be, very
revealing, for two main reasons:
firstly, in this band, we have the contributions of both the radiation
processes (synchrotron and inverse Compton) thought to originate 
the overall continuum of blazars.
In this respect {\it Beppo}SAX with its 0.1--100 keV band
has been particularly revealing. 
Secondly, the superior angular resolution of the {\it Chandra} satellite 
is giving us pictures of the large scale jets, which, somewhat
unexpectedly, are very bright in X--rays.

We are now beginning to construct a coherent picture of the 
blazar phenomenon, which is important not only {\it per se} but 
especially to understand how the relativistic jets work and what
originates them.
Their emitted radiation is and must be (in order to power the 
extended radio structures) only a small fraction (a few per cent) 
of the power they carry.
In the most powerful jet sources, the estimated limits on the
jet power rivals the power that can be extracted by accretion.
The machine which forms and powers the jets in radio sources
is therefore as important as accretion: this is why the study of 
blazars is important.

The $\gamma$--ray radiation we see is intense and variable.
This suffices to constrain where this radiation is produced:
it cannot be a region too compact or too close to the accretion disk 
and its X--ray corona, to let $\gamma$--rays survive against 
the $\gamma$--$\gamma \to e^{\pm}$ absorption process,
and it cannot be too large in order to vary rapidly (Ghisellini 
\& Madau 1996).

Furthermore, we know that the Spectral Energy Distribution
(SED) of blazars is always characterized by two broad peaks,
thought to be produced by the two main radiation processes,
i.e. synchrotron at low frequencies and inverse Compton at
high energies (see e.g. Sikora 1994 for a review,
but see e.g. Mannheim 1993 for a different view).
The relative importance of the two peaks (i.e. processes)
and their location in frequency appear to be a function of the 
total power of blazars (Fossati et al., 1998, Ghisellini et al., 1998),
leading to a {\it blazar sequence}.
Giommi \& Padovani (1994) were the first to notice that
the different flavors of blazars corresponded to a different
location of their synchrotron peak, and called LBL and HBL
the BL Lac objects having the synchrotron peak at low or 
high frequency, respectively, and I will use in the following 
the more ``colorful" division in red, green and blue blazars
(proposed by L. Maraschi) where the color obviously refers to
the frequency of the peaks.

\section{Red blazars}

These powerful blazars include flat spectrum radio quasars 
with relatively strong emission lines, and several BL Lacs
of the LBL--type. 
The synchrotron peak is in the mm--far IR band, while the high energy
peak is in the MeV band, and is largely dominating the power output.
In Fig. 1 we show two ``extremely red" sources, 1428+4217 and 0836+710, 
which are among the most powerful sources known (if isotropic, the 
emitted power would exceed $10^{49}$ erg s$^{-1}$).
These blazars are important because these are the second most powerful
engines to produce bulk kinetic energy, after Gamma--Ray Bursts. 
By dividing the observed luminosity by the square of the Lorentz
factor we can estimate a lower limit on the jet power,
assuming that the jet carries more power than what it
produces in radiation (i.e. that the emission process is more or less
continuous, with no accumulation and rapid release of random energy).
In this way we derive jet powers for these surces of the
order of $P_{\rm jet} \sim 10^{47} L_{49}/\Gamma_1^2$ erg s$^{-1}$.
\begin{figure}
\plottwo{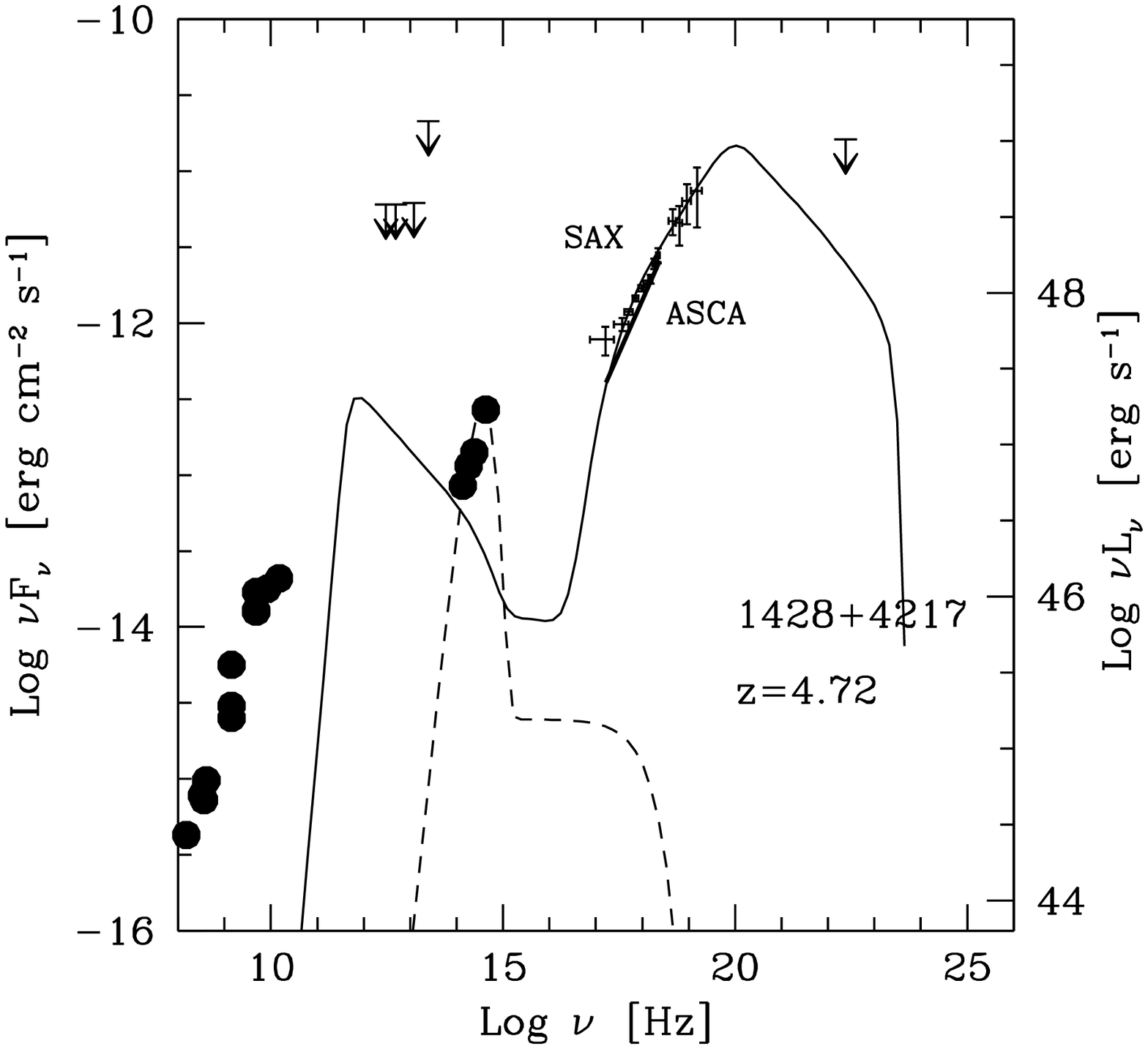}{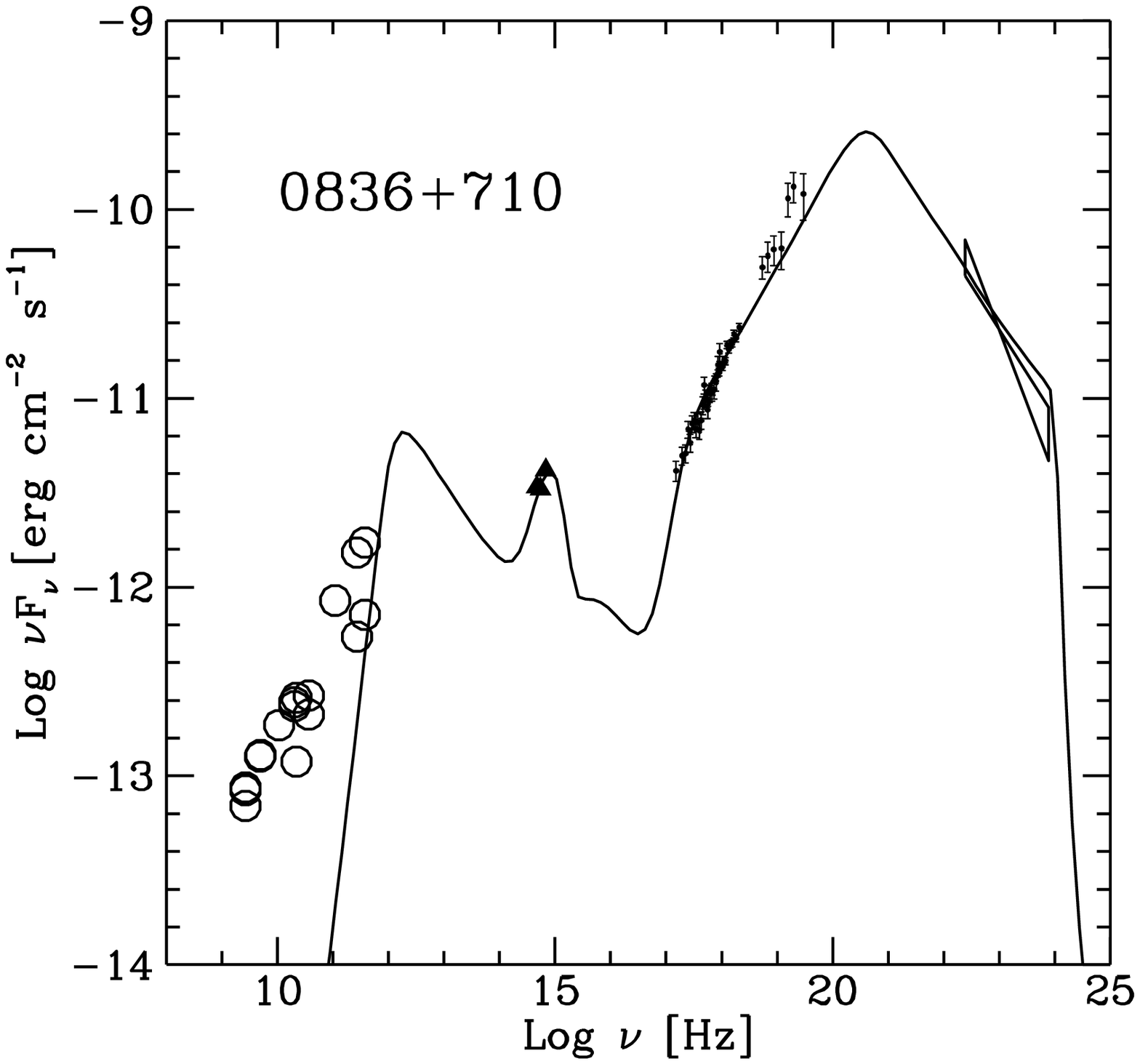}
\caption{Examples of red blazars.
{\bf Left:} SED of 1428+4217 with the recent {\it Beppo}SAX
observations, revealing that the X--ray spectrum, likely
due to the invese Compton process, is dominating the power output.
Note that this is the most distant radio--loud quasar known.
The solid line refers to a one zone homogeneous synchrotron
inverse Compton model.
Adapted from Fabian et al., 2000.
{\bf Right:} SED of PKS 0836+710. Again, the power output in this 
$z=2.17$ blazar is dominated by the high energy emission, as revealed by 
{\it Beppo}SAX and EGRET. From Tavecchio et al., 2000. }
\end{figure}


Red blazars are not very conspicuous in the optical, since 
their synchrotron peak is at lower frequencies, and they 
are the sources with the largest X--ray to optical flux ratio.
The two sources shown in Fig. 1 surely have their jets pointing at us,
and therefore their radiation is maximally boosted by beaming.
Other blazars should exist with somewhat larger viewing angles 
(but not too large, not to appear as a radio--galaxy) which are less powerful 
and even redder.

\section{Green blazars}
\begin{figure}
\plottwo{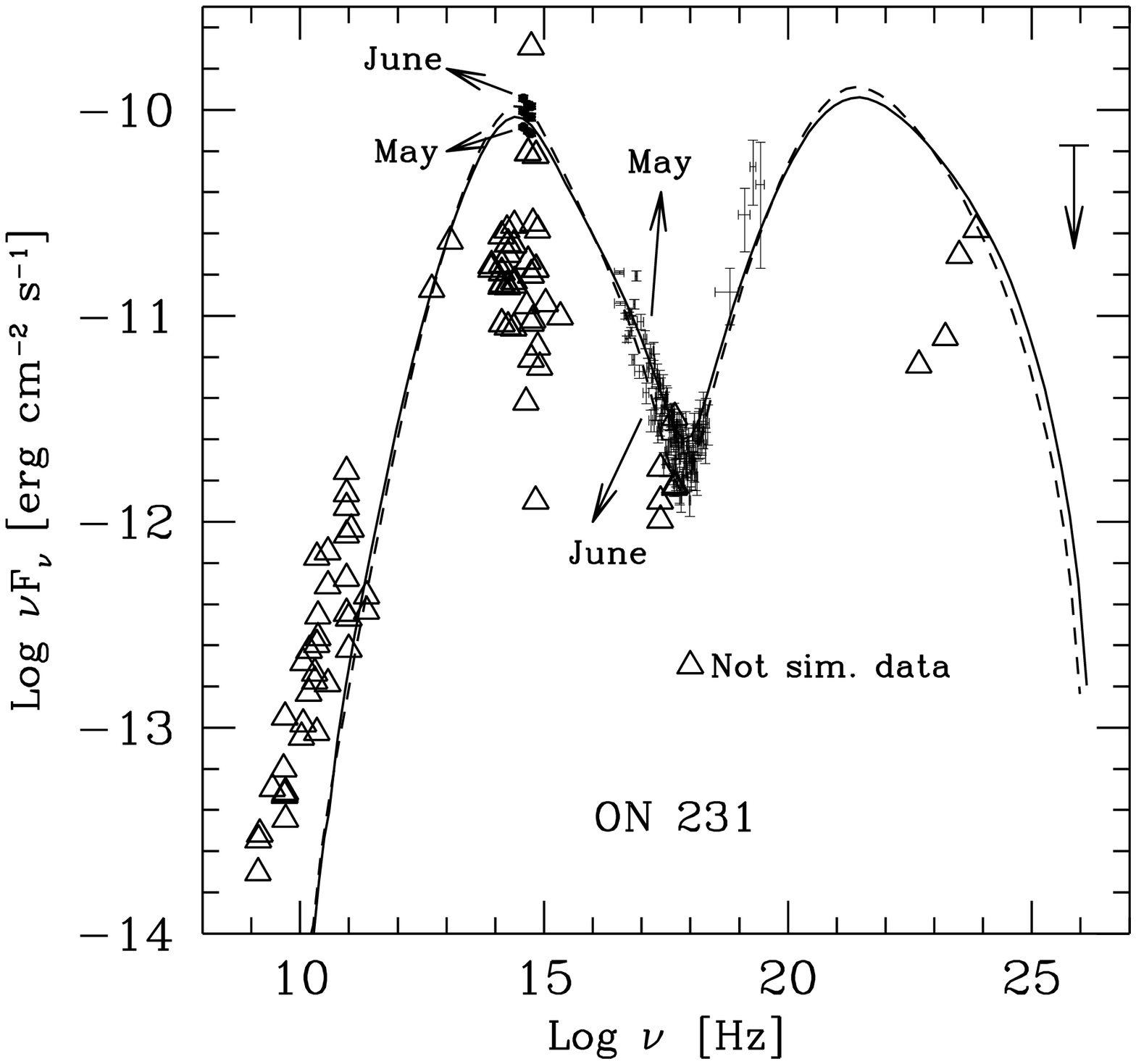}{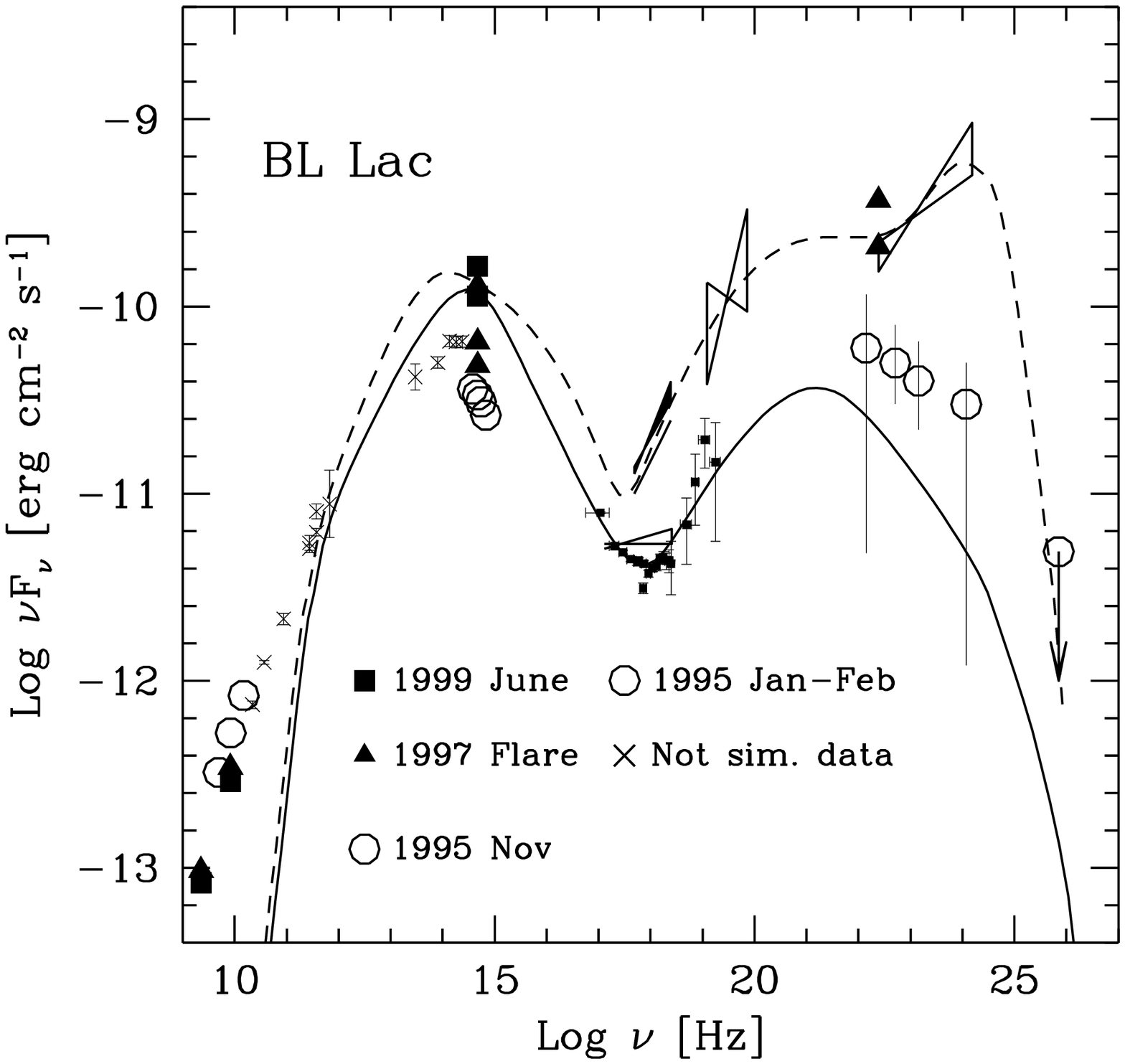}
\caption{Examples of green blazars.
{\bf Left:} SED of ON231 with the recent {\it Beppo}SAX
observations, revealing that the X--ray spectrum
is nicely described by the high energy tail of the synchrotron
at lower energies, while above a few keV, the self Compton process
strongly hardens the spectrum.
From Tagliaferri et al., 2000.
{\bf Right:} SED of BL Lac in the flaring state of summer 1997 and
in the more recent {\it Beppo}SAX observation in 1999.
During the latter observation, the optical flux was as high
as during  the 1997 flare, but the X--ray emission was much lower,
revealing both the synchrotron and the inverse Compton components.
From Tagliaferri et al., 2000.
}
\end{figure}
Fig. 2 shows two examples of ``green" blazars, ON 231 and BL Lac, 
which are blazars with intermediate properties.
Most of them are classified as LBL blazars,
and often their X--ray spectrum shows  both
a steep power law energy component (identified as
the high energy tail of the synchrotron emission) and
a very hard high energy power law (identified
as the emerging of the inverse Compton component).
During the {\it Beppo}SAX observations of BL Lac, the
source had a very rapid variability (timescales of 20 minutes)
in the soft energy band, absent at high energies.
We expect this behavior to be typical in this class of sources,
and could be due to different emitting zones or to the different 
cooling times of the radiating electron (the most energetic are
producing the synchrotron tail while the hard X--rays
are produced by electrons with much less energy and longer cooling times).

\section{Blue blazars}

\begin{figure}
\plottwo{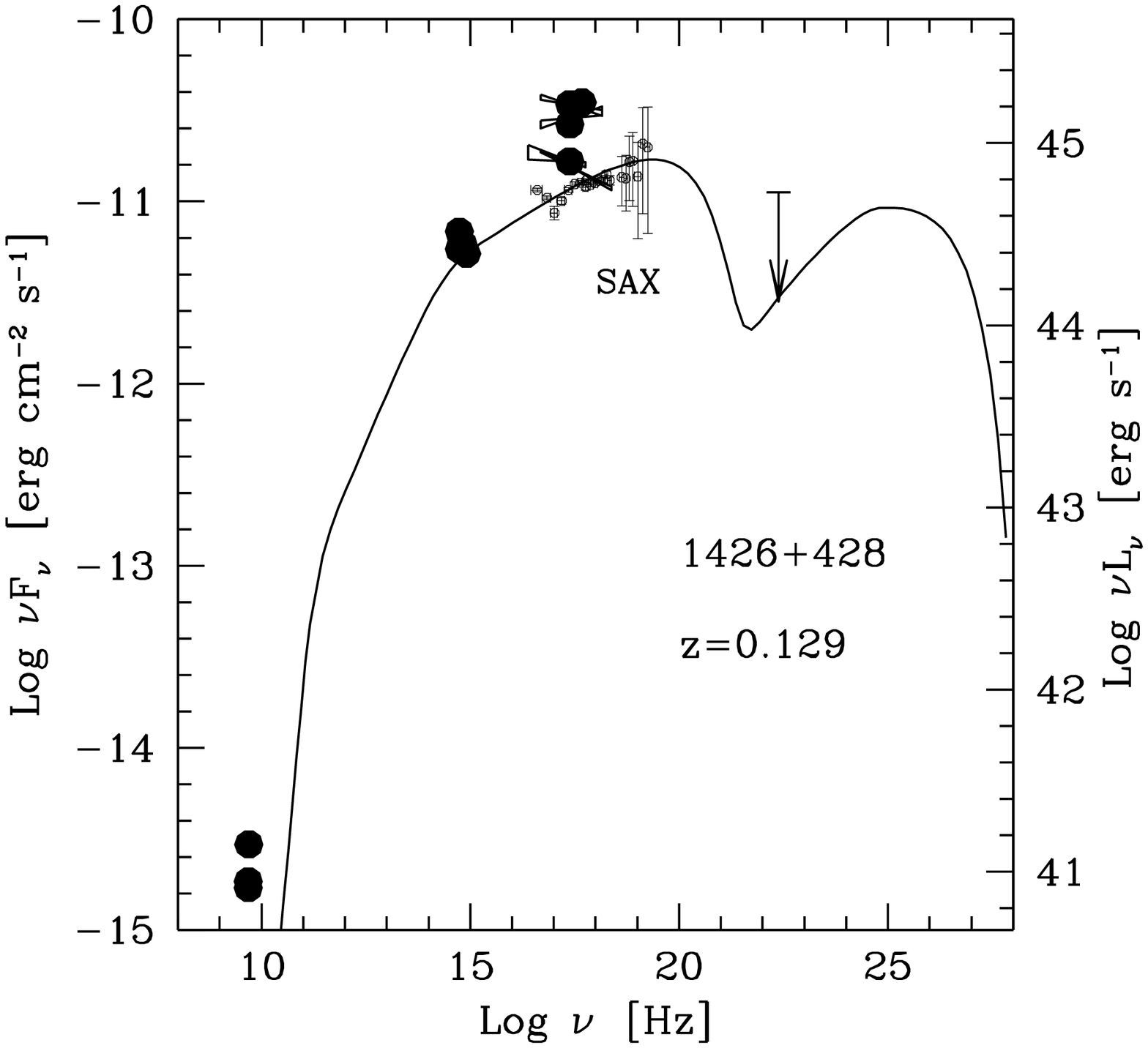}{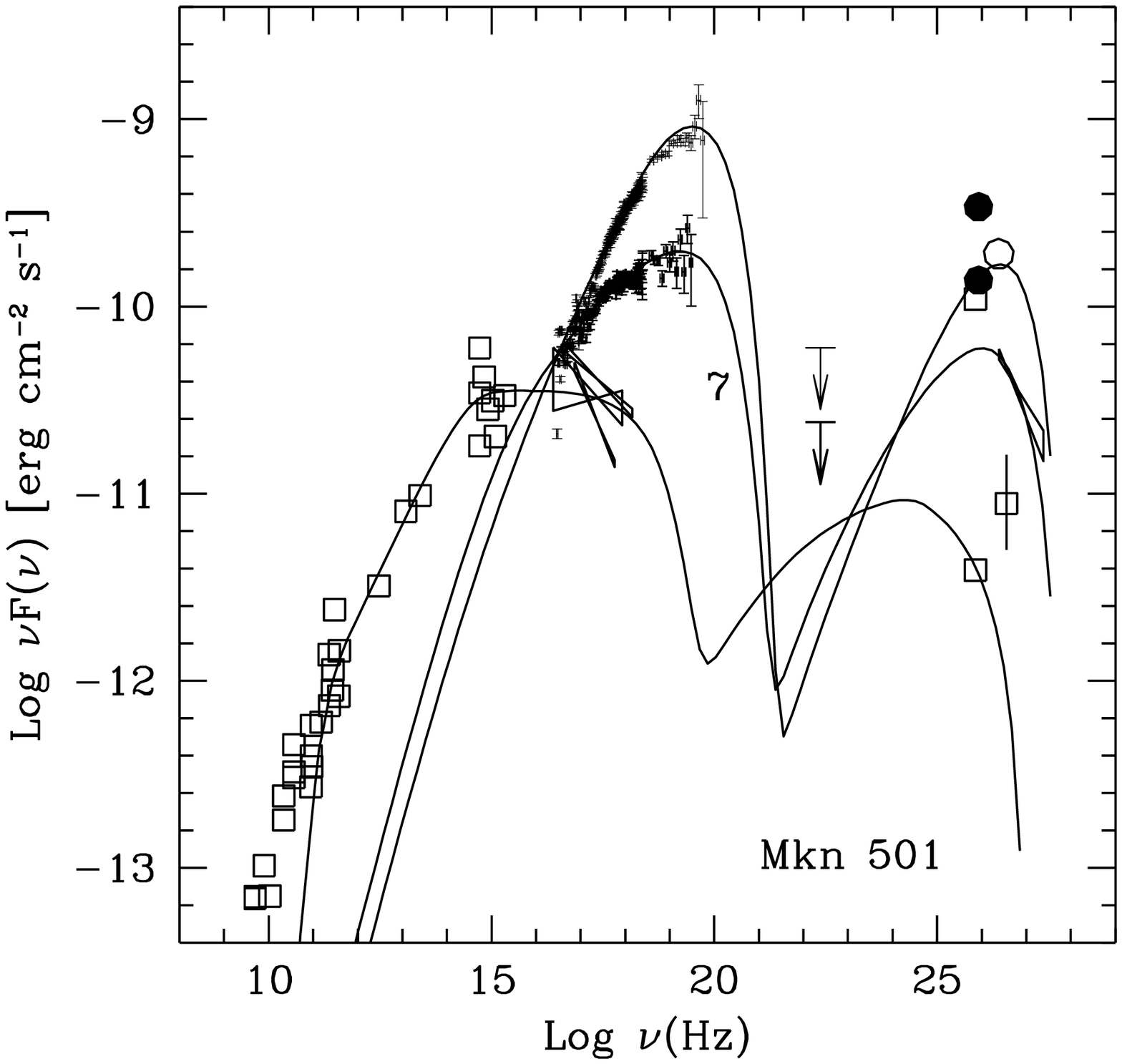}
\vskip -0.3 true cm
\caption{Examples of blue blazars.
{\bf Left:} SED of 1ES 1426+428 with the recent {\it Beppo}SAX
observations, revealing a synchrotron peak located at or above 
100 keV. 
The solid line refers to a one zone homogeneous SSC model.
Adapted from Costamante et al., 2000.
{\bf Right:} SED of Mkn 501 during a quiescent state and during the
flaring state of 1997. Adapted from Pian et al., 1998.
Solid lines are SSC models as discussed in Ghisellini 1998.
}
\end{figure}

These are the less powerful blazars, with the synchrotron peak located
above the optical band, and sometimes reaching the hard X--ray band, 
as in 1ES 1426+428 and Mkn 501, whose SEDs are shown in Fig. 3.

High energy (especially in the X--ray and TeV bands) observations are
important in these sources because there we see the radiation produced
by the most energetic electrons (with random Lorentz factors $\gamma>10^6$),
which can then reveal some properties of the acceleration mechanism 
at its extreme.
Furthermore, these sources are the best candidates to emit copiously in the 
TeV band, as in the case of Mkn 421, Mkn 501, PKS 2155--304 and 1ES 2344+514.
Besides giving information on the emission and acceleration mechanism, 
TeV--band data can also allow the measurement of the amount of absorption 
(through the $\gamma$--$\gamma \to e^{\pm}$ process) both local to the source
and the one due to the infrared backgroung radiation 
(see e.g. Stecker \& De Jager, 1997)

One can ask if sources even more extreme than Mkn 501 and 
1ES 1426+428 can exist, with the synchrotron peak in the MeV range.
Shock theory does not prevent it, since the intrinsic maximum
predicted synchrotron energy is $\sim$70 MeV.
According to our proposed sequence, these sources
should be low power objects and modest in all bands but the MeV one,
since they should be radio--weak and dominated by the light of
the galaxy in the optical.
Even in the ``usual" 2--10 keV X--ray band their synchrotron emission, 
while rising, would still be weak.
The self Compton emission in the TeV (and beyond) could be intrinsically
faint because of Klein Nishina effects.
Therefore there may be many of these sources, hidden in normal 
elliptical galaxy, with jets of overall low power pointing at us, 
emitting most of their emission in the MeV band.
Future INTEGRAL (in the MeV band) and VERITAS (in the TeV band) 
observations seem the most promising to discover these 
``synchrotron MeV BL Lacs" (or UV blazars).

\section{What controls the SED of blazars?}

In the previous sections we have seen some examples
of the variety of the SEDs of blazars.
From these examples there seems to be a link between 
the ``color" of a blazar (i.e. the location of its peaks)
and its overall luminosity.
This impression has been confirmed by a much more detailed study
by Fossati et al. (1998), who considered the largest complete
blazar samples known at that time, and divided the sources with
respect to their radio luminosity (thought to trace well the bolometric 
power).

As a result we found that blazars come in a sequence, whose main
parameter is the observed power:
with increasing power blazars tend to be redder, with
a more dominating high energy peak.
At first sight this is surprising, since the {\it observed} power
is enhanced by beaming by orders of magnitude, and
changes in viewing angle and/or bulk Lorentz factor can
dramatically change the observed flux, hence the power.
If the beaming factor is $\delta$, the observed bolometric
flux is $\propto \delta^4$ while the peak frequencies
is $\propto \delta$: note that beaming tends to give
a correlation {\it opposite} to the one observed:
more powerful object should be bluer, not redder.
Therefore one interesting conclusion is that
blazars are characterized by the same degree of beaming,
i.e. the bulk Lorentz factors are contained in a narrow range.
The viewing angles should be similar in all blazars as well, 
and this counter--intuitive conclusion may suggest
that the emitting plasma in jets is moving in a fan, not 
always parallel to the jet axis, and we perhaps call blazars 
those sources observed {\it within} the jet opening angle.

Ghisellini et al. (1998) applied to all blazars detected by EGRET 
(with known distance and $\gamma$--ray spectral shape),
a simple, one--zone, homogeneous synchrotron inverse Compton model 
including the possible contribution to the inverse Compton process
of photons produced externally to the jet.
This enabled us to estimate the intrinsic parameters of the source,
such as the magnetic field, the radiation energy density, the size, 
the beaming factor and the energy of the electrons emitting at the two 
peaks of the SED, or, equivalently, their random Lorentz
factor $\gamma_{\rm peak}$.
We found a remarkable correlation between $\gamma_{\rm peak}$
and the total energy density $U$ (radiative plus magnetic)
as seen in the comoving frame (see Fig. 4, left panel).
The slope of this correlation $\gamma_{\rm peak}\propto U^{-0.6}$
strongly suggests that {\it radiative cooling} is playing a crucial role
in determining this correlation.
The radiative cooling rate scales as $\dot\gamma \propto U\gamma^2$
and the found correlation slope implies that, at $\gamma_{\rm peak}$,
all sources have the same cooling rate.

The model we used for the fitting assumed that there is a continuous 
injection of relativistic particles throughout the source, at a rate 
$Q(\gamma) \propto \gamma^{-s}$ above some random Lorentz factor 
$\gamma_{\rm min}$.
We then assumed that radiative cooling was dominating at all 
energies, resulting in a steady particle distribution characterized 
(in the absence of Klein N ishina effects) by a broken power law, 
being $\propto \gamma^{-2}$ at low energies, up to $\gamma_{\rm min}$,
and $\propto \gamma^{-(s+1)}$ above.
In this case (since $s$ was larger than 2), the peaks of the spectrum
are due to electrons with $\gamma_{\rm peak}=\gamma_{\rm min}$.
\begin{figure}
\plottwo{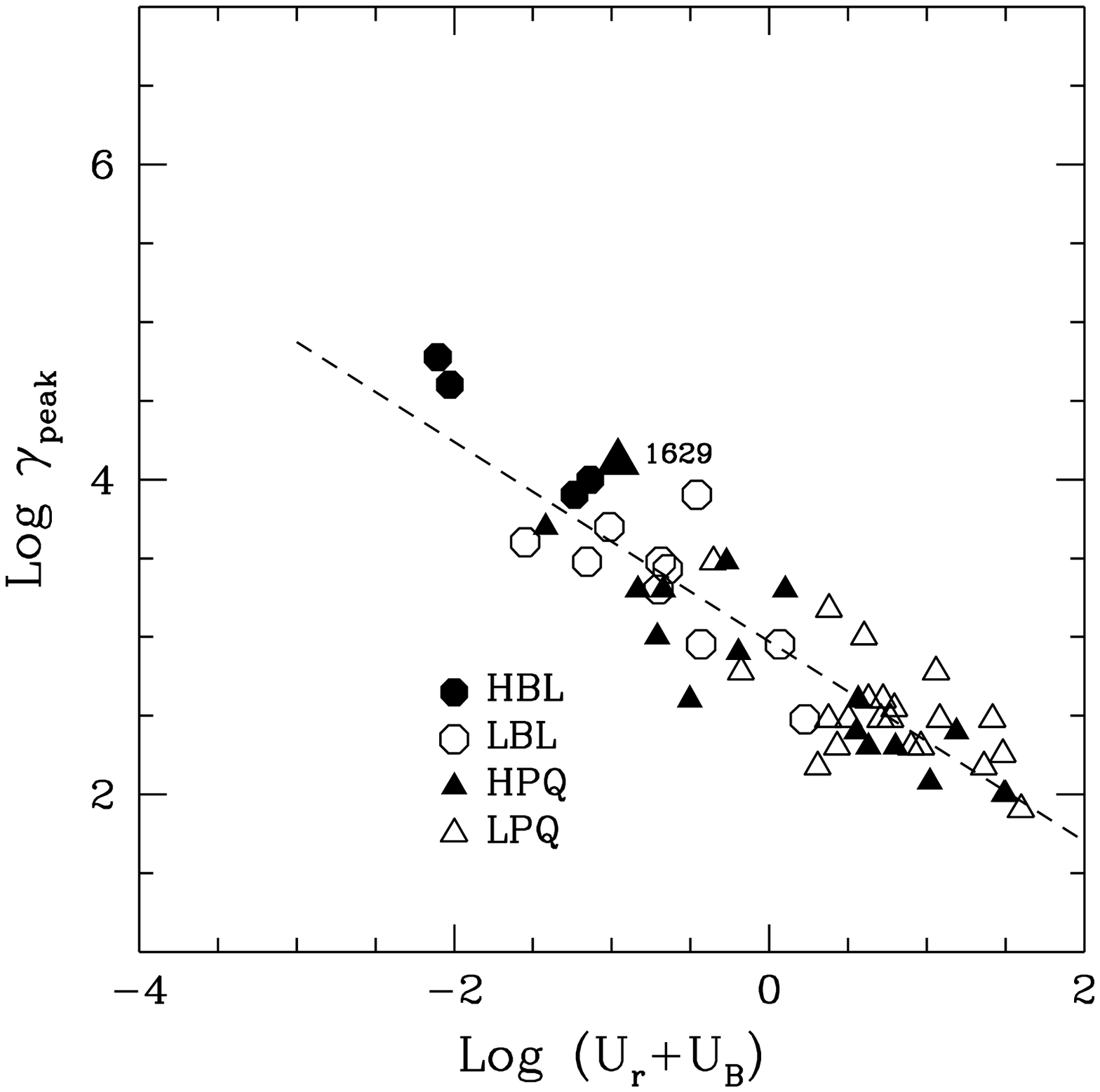}{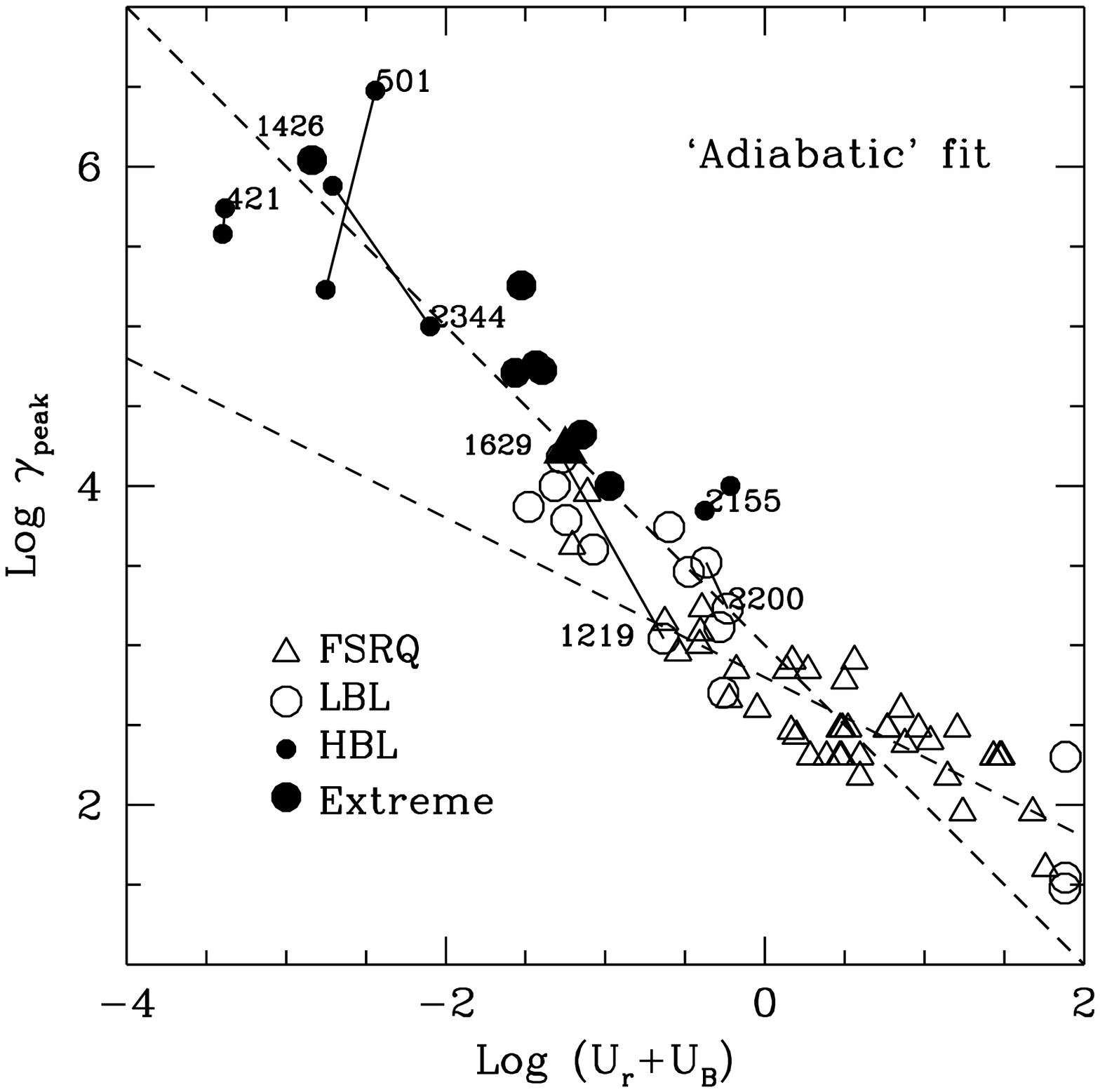}
\vskip -0.3 true cm
\caption{
{\bf Left:} The random Lorentz factors of the electrons emitting most
of the blazar radiation, $\gamma_{\rm peak}$ as a function of the 
intrinsic energy density (magnetic plus radiative, including external photons).
In this case $\gamma_{\rm peak}$ corresponds to the less energetic
electrons assumed to be continuosly injected throughout the source.
The correlation is of the form $\gamma_{\rm peak}\propto U^{-0.6}$.
Note that only a few HBL are present. From Ghisellini et al., 1998.
{\bf Right:} Same as in the left panel, but now including more 
HBL and extreme BL Lacs, and including adiabatic losses in the computation
of the particle energy distribution.
In this case $\gamma_{\rm peak}$ is the same as before 
in powerful blazars where the radiative cooling is complete,
but becomes larger when adiabatic losses are important.
Note the two power law branches of the correlations.
From Ghisellini \& Celotti in prep.
}
\end{figure}

More recently, we (Ghisellini \& Celotti, in prep.) have investigated 
in more detail this issue, testing if this correlation holds also 
for more extreme sources, i.e. extremely blue blazars.
Many sources of this class have been observed recently by {\it Beppo}SAX,
and for them we can construct meaningful SED even if we ignore the location
of the high energy peak (these sources have not been detected by EGRET,
nor by Cherenkov telescopes).

Despite the uncertainties introduced by the unknown level of the Compton
emission, the intrinsic parameters resulting from the fits
are limited in a given range {\it which is not compatible with
the $\gamma_{\rm peak}\propto U^{-0.6}$ relation}.
For these extreme sources the relation is steeper 
(i.e. $\gamma_{\rm peak}\propto U^{-1}$, see Fig. 4, right panel).

We then fitted the SED with a slightly different model,
in which the particle energy distribution takes into account
the relative importance of adiabatic vs radiative cooling.
We then find that for powerful blazars the radiative cooling
is always dominating (and then find the same results as above),
but for blue blazars adiabatic can dominate over radiative cooling
even at energies $\gamma>\gamma_{\rm min}$.
In this case the peak of the emission is produced by electrons
for which the two cooling processes are equal.
This effect should be responsible of the  
$\gamma_{\rm peak} \propto U^{-1}$ relation we find for blue blazars
(Ghisellini \& Celotti in prep).

\section{Large scale X--ray jets}

{\it Chandra} is finding that jets of both blazars and radiogalaxies 
show X--ray emission at large scales (Chartas et al. 2000;
Schwartz et al. 2000; Wilson et al., 2000).
In the case of the blazar PKS 0637--752 [a superluminal source (Lovell 2000),
hence observed under a very small viewing angle], the X--ray flux comes
from a (deprojected) distance of $\sim$1 Mpc from the center.
These observations imply the presence of a large number of
relativistic electrons, and the relatively faint optical
emission implies that the X--rays cannot be produced by the 
synchrotron process (see Fig. 5).

We (Celotti, Ghisellini \& Chiaberge 2000) have proposed 
that the X--ray flux in PKS 0637--752 is inverse Compton 
scattering with the cosmic background radiation, 
which in the comoving frame of the source is seen enhanced 
by blueshift and time contraction.
We then require that, at the Mpc scale, 
the jet is still relativistic $\Gamma=10$--15) and therefore
that the produced X--ray rays are beamed.

Although it may seem odd to have relativistic motion at such large 
distances, consider that {\it it is the most economic} way to 
account for the observed X--ray power.
In fact for lower values of $\Gamma$ more emitting 
electrons are needed, and the total bulk kinetic power carried 
by them {\it increases} despite the smaller $\Gamma$--factor.
\begin{figure}
\vskip -0.3 true cm
\plottwo{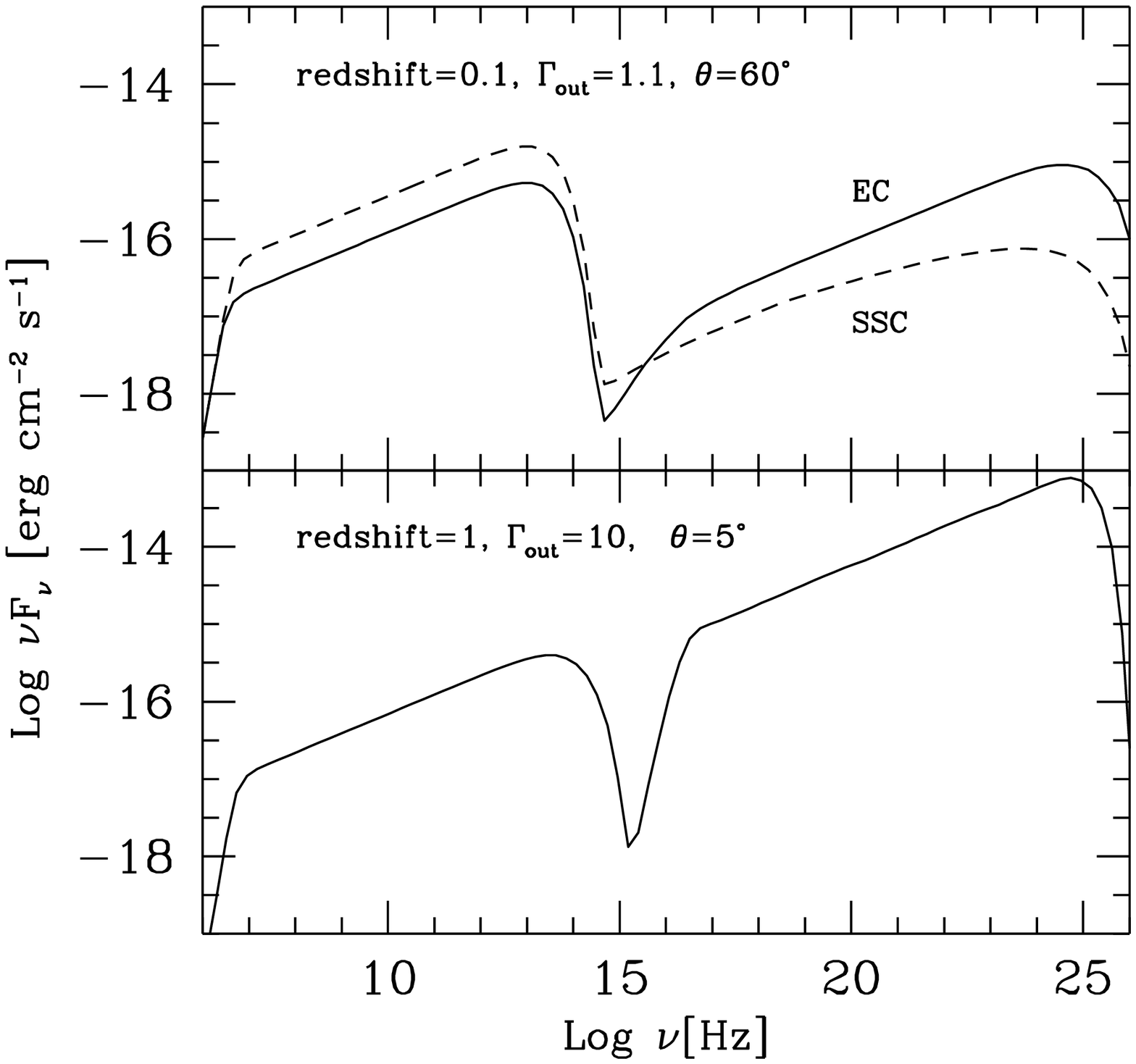}{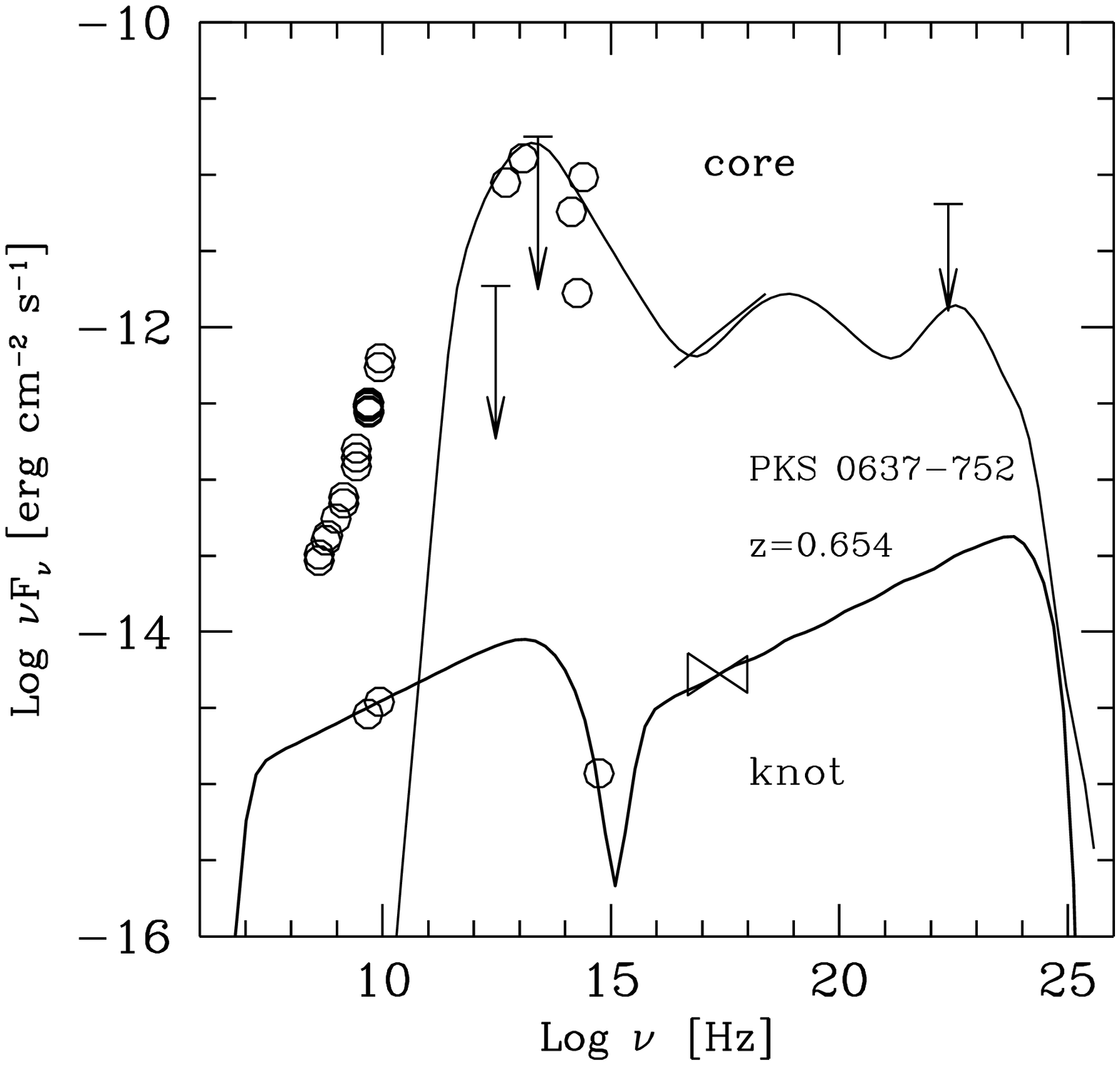}
\vskip -0.5 true cm
\caption{Examples of spectra expected form large scale jets.
{\bf Left:} At 10 kpc from the center a region of 1 kpc of size 
embedded in magnetic field of $10^{-5}$ G radiates an intrinsic 
power of $3\times 10^{41}$ erg s$^{-1}$.
The nuclear (blazar) component emits an intrinsic power of $10^{43}$ 
erg s$^{-1}$. 
The {\it upper panel} shows the emission from a layer with 
$\Gamma_{\rm layer}=1.1$ and viewing angle $\theta=60^\circ$.
The dashed line corresponds to the SED assuming SSC radiation only.
The solid line accounts for radiation coming from the 
core of the jet and illuminating the region.
The {\it bottom panel} shows the emission from a spine moving with
$\Gamma=10$ at a viewing angle $\theta=5^\circ$.
{\bf Right:} The SED of the core and the large scale knot of PKS 0637--752,
together with the models for both components (solid lines). From 
Celotti, Ghisellini \& Chiaberge, (2000).
}
\vskip -0.3 true cm
\end{figure}
We also pointed out that the large scale jet could have some
velocity structure (Celotti et al., 2000; Chiaberge et al., 2000),
consisting in fast ``spines"  and slowly moving ``layers".
The layers could consist of dissipation zones of the jet where
the jet material collides with obstacles in the jet or its walls:
particles can be accelerated there and produce quasi--isotropic
synchrotron emission.
If these regions are not misaligned with respect to the inner (sub--pc)
jet, they can be illuminated by the intense and beamed radiation
coming from the sub--pc core of the jet, and hence receive an extra 
contribution of seed photons to be Compton scattered at high energies, 
enhancing the X--ray emission.

Fig. 5 (left, top panel) shows one example of enhanced Compton emission
with respect to a pure SSC spectrum.
Being slow, and possibly only mildly relativistic, the layers produce 
radiation which is much more isotropic than the spine: at small
viewing angles (blazar case) this component is outshined by
the spine component, but it can become visible as the viewing
angle increases (i.e. in radio--galaxies).

\section{Jet power}

\begin{figure}
\plotone{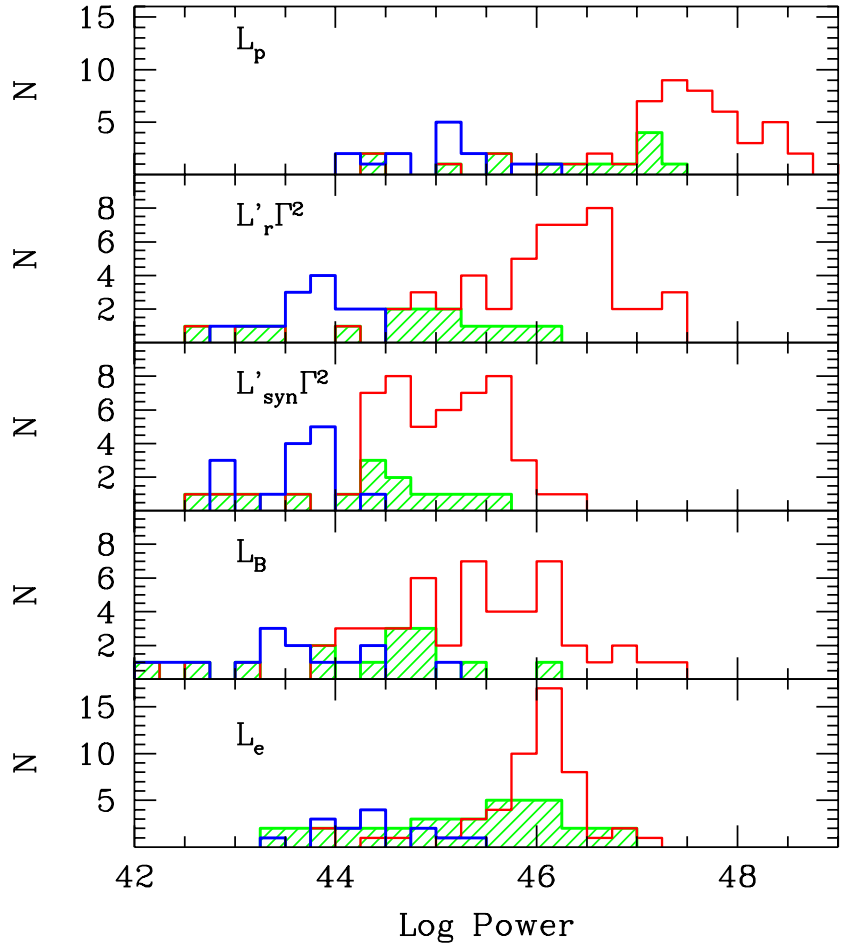}
\caption{Histograms of the distribution of kinetic power
(in erg s$^{-1}$)
of the blazars considered in Celotti \& Ghisellini 2001.
$L_{\rm p}$ is the power carried by protons assuming one
proton per emitting electron;
$L^\prime_{\rm r}\Gamma^2$ is the power radiated by the jet
and $L^\prime_{\rm syn}\Gamma^2$ the power in synchrotron
radiation only;
$L_{\rm B}$ is the Poynting flux;
$L_{\rm e}$ is the power carried by the emitting electrons.
The gray thin histograms and the shaded areas correspond to FSRQs and BL Lacs
respectively, detected by EGRET and with
measured $\gamma$--ray spectrum.
The bold line histograms correspond to extreme BL Lacs
not detected by EGRET, for which we have assumed that the SSC
emission has about the same power as the synchrotron one.
}
\end{figure}

Due to beaming, it is not trivial to estimate the amount of
luminosity intrinsically emitted by blazars and radio--galaxies, 
and then to use it to estimate the power carried by jets.
Powerful FR II radiogalaxies and quasars, on the other hand,
have extended radio structures and lobes containing huge
quantity of energy, which we can estimate resorting to 
equipartition arguments and simplifying assumptions.
These structures can be thought of as calorimeters,
since there the radiative cooling time of the emitting particles 
is long: by knowing the lifetimes of these structures we can then 
estimate the average incoming power that jets must provide.
This has been done, among others, by Rawlings \& Saunders (1991)
who also found a correlation between the average power required 
by radio lobes and the luminosity of the narrow lines.
These power estimates are appropriate for the largest scales, 
i.e. 100--1000 kpc.

Celotti \& Fabian (1993) and Celotti, Padovani \& Ghisellini (1997)
have instead directly estimated the 
power carried by jets by estimating the number of electrons
required to account for the radio flux seen in VLBI components,
hence at the 1--10 pc scale.

Finally, Celotti \& Ghisellini (2001, in prep) estimate the
jet power by computing the number of particles required to 
account for the bulk of the observed luminosity (often
coinciding with the $\gamma$--ray luminosity) by applying
a simple one--zone homogeneous synchrotron inverse Compton model 
to the SED of blazars.
These estimates refer to the inner part of the jets, i.e. to the
0.01-0.1 pc scale.
The distributions of the bulk kinetic
power carried by jets in the form of protons ($L_{\rm p}$,
assuming one proton for each emitting electron);
emitted total radiation ($\Gamma^2 L^\prime_{\rm r}$, where
$L^\prime_{\rm r}$ is the emitted power as measured in the comoving frame);
emitted synchrotron radiation ($\Gamma^2 L^\prime_{\rm syn}$);
Poynting flux ($L_{\rm B}$) and electrons ($L_{\rm e}$) are shown in Fig. 6.
As can be seen, the dissipated power ($\Gamma^2L^\prime_{\rm r}$) is greater,
on average, than the power carried by electrons and magnetic field only, 
(in this case the jet would dissipate more than what it can),
implying an energetically important proton component.
In the absence of electron--positron pairs we would have the
$L_{\rm p}$ distribution shown in Fig. 6, whose average value
is a factor 10 larger then the average dissipated power.
Considering that most of the jet power must be transported out
to reach the extended radio structures, there is little room
for a large amount of electron--positron pairs (if present, they
would decrease $L_{\rm p}$ because there would be less than one
proton for each emitting lepton).

The histograms in Fig. 6 assumed that the particle distribution of
emitting electrons had no low energy cutoff, i.e. $\gamma_{\rm min}=1$.
This parameter is important because the particle number 
depends on its value, which in turn is very poorly constrained by
observations.
On the other hand the plotted $\Gamma^2 L^\prime_{\rm r}$
distribution {\it does not depend on it.}
We therefore conclude that, in order for the bulk kinetic
jet power to be larger than the dissipated power, $\gamma_{\rm min}$
cannot be larger than a few (in other words: many particles are needed
to transport a power larger than what is wasted).

\section{Internal shocks}

\begin{figure}
\plottwo{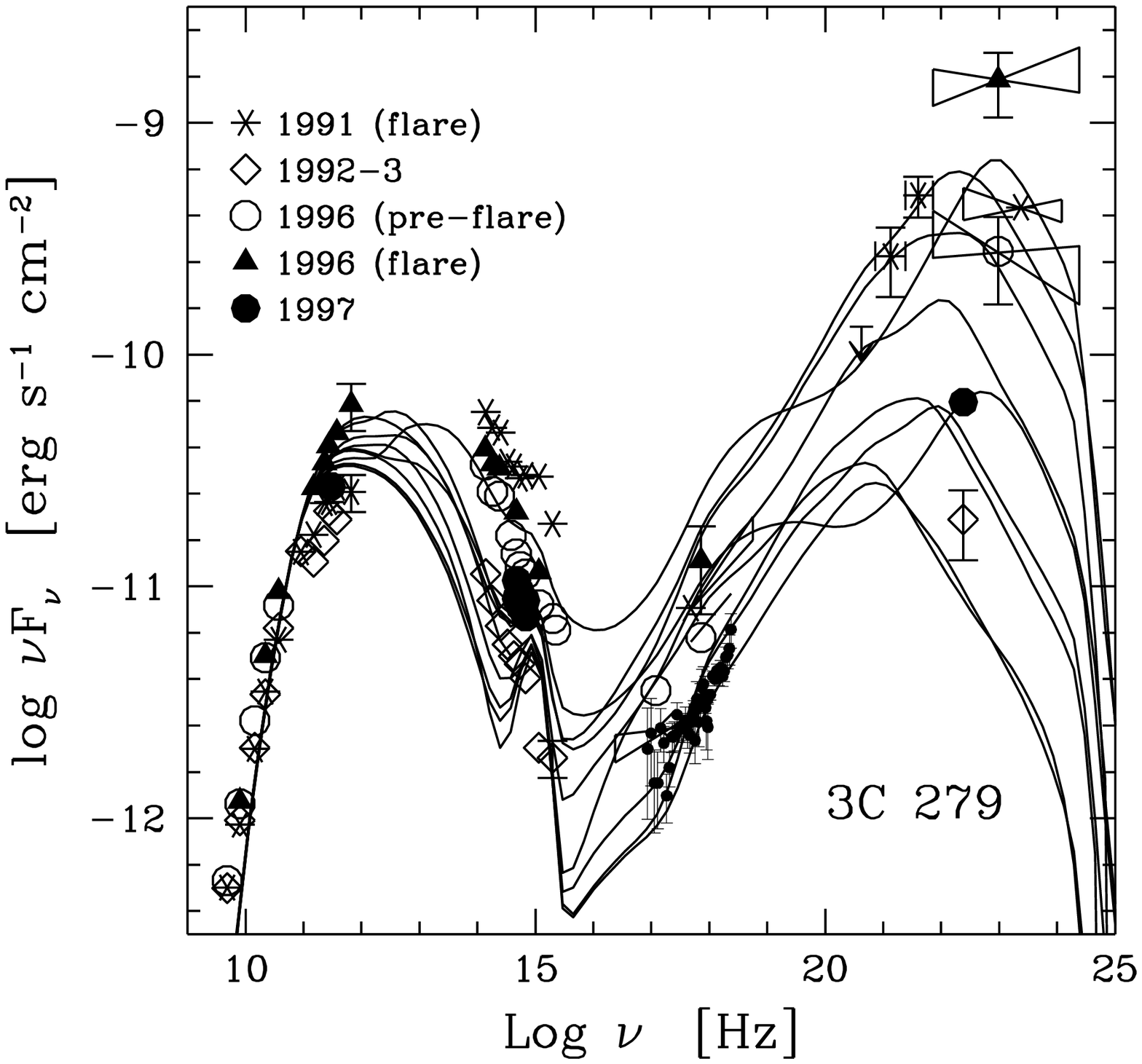}{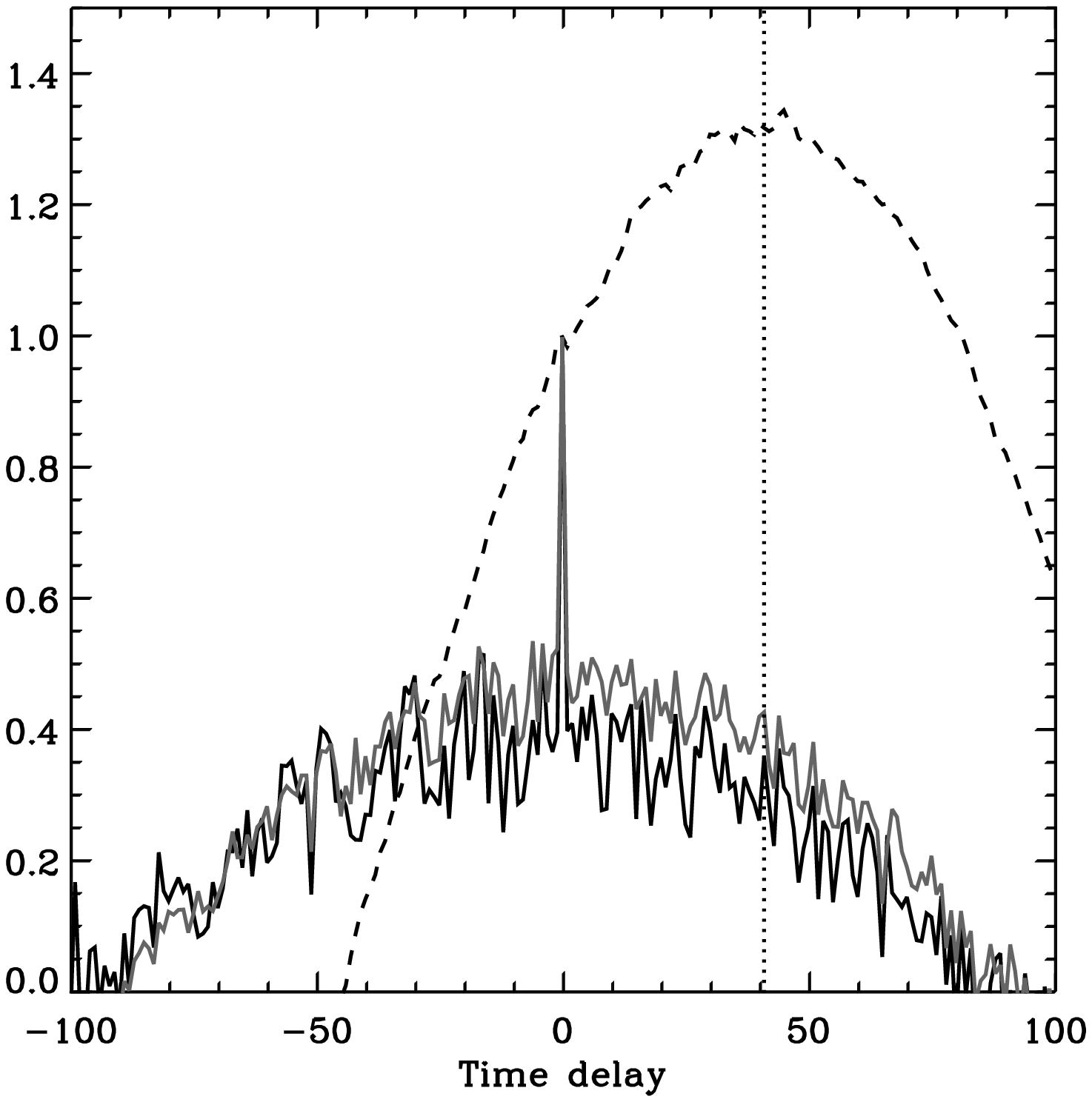}
\caption{
{\bf Left:} SED of 3C 279 with, superimposed, different
spectra resulting from our simulations of internal shocks.
These correspond to different shell--shell
collisions in different parts of the jet.
{\bf Right:} Cross correlation between the simulated $\gamma$--ray
light curve and the X--ray, the optical and the mm light curves.
Only for the latter there is a clear delay of about 40 days,
due to the average distance of the regions producing
$\gamma$--rays and mm radiation, once Doppler time contraction
is accounted for.
From Spada et al., 2001.
}
\end{figure}
In gamma--ray burst science, the most accepted scenario 
for explaining the origin of the prompt emission is the so called
{\it internal shock scenario}, in which the central engine
works intermittently,
producing shells of slightly different velocities,
mass and energy (e.g. Rees \& M\'esz\'aros 1994).
Faster and later shells can then catch up slower earlier ones,
dissipating part of their bulk kinetic energy into radiation.
This model could work even better in blazars: for them we require
a {\it moderate} efficiency for the bulk to random energy
conversion, as is the case in this scenario.
Indeed, this idea was born in the blazar field (Rees 1978),
and only later became the leading idea to explain the gamma--ray burst 
emission (but see Sikora, Begelman \& Rees 1994).
This model is very promising, since it can explain some basic properties
of blazars:

\begin{itemize}

\item The efficiency is of the right order: most of the jet power
{\it has not to be dissipated}, in order to power the radio lobes.

\item If the initial separation is comparable to a few Schwarzchild radii,
i.e. $R_0\sim 10^{15}$ cm, the collision takes place at $R_i\sim 10^{17}$ cm
(for $\Gamma_1\sim 10$), just at the distance where the inverse
Compton scattering off the photons of the Broad Line Region is efficient,
where  the $\gamma$--$\gamma to e^\pm$ process is not important, and yet the
emitting region is still sufficiently compact to account for the
rapid variability.

\item There can be a hierarchical structure in shell--shell collisions:
pairs of shells can collide once more, at greater distances, where the 
dominant channel for radiation is synchrotron emission with a smaller
value of the magnetic field.
Hence there can be a link between the flares at optical and $\gamma$--ray
energies and the flares in the radio--mm band (see Fig. 7, right panel).

\end{itemize}

These qualitative properties have been verified by numerical simulations
by Spada et al. (2001), assuming a jet of average bulk kinetic power of 
$10^{48}$ erg s$^{-1}$, carried by shells or blobs injected in the jet, 
on average, every few hours, with a bulk Lorentz factor chosen at random 
in the range [10--30].
The first collisions happen at a few$\times 10^{16}$ cm, well within
the Broad Line Region (BLR), assumed to be located at $5\times 10^{17}$ cm
and to reprocess 10\% of a disk luminosity, of the order of  
$10^{46}$ erg s$^{-1}$.
Particles emit by synchrotron,
 synchrotron self--Compton 
and Compton scattering off the external radiation (EC) produced by the BLR. 
In Fig. 7 (left panel) we show some spectra, each corresponding to one 
single shell--shell collision at a different distance, and 
the entire time dependent evolution 
can be seen in the form of a movie at the URL:
{\tt http://www.merate.mi.astro.it/$\sim$lazzati/3C279/index.html}.


\vskip 0.2 true cm
{\bf Acknowledgments}: I thank Annalisa Celotti for years of fruitful
collaboration.

\end{document}